\documentclass[showpacs,aps,amsmath,amssymb,twocolumn]{revtex4}

\usepackage{graphicx}
\usepackage{dcolumn}
\usepackage{bm}
\usepackage{graphicx}
\usepackage{graphics}
\usepackage{amsmath}
\begin{document}

\title{Landau Quantization of Neutral Particles in  an External Field}
\author{ Claudio Furtado$^{1}$\footnote{Electronic Address:
furtado@fisica.ufpb.br}, J. R.  Nascimento and  L. R. Ribeiro }

\affiliation{$^{1}$Departamento de F\'{\i}sica, CCEN,  Universidade Federal
da Para\'{\i}ba, Cidade Universit\'{a}ria, 58051-970 Jo\~ao Pessoa, PB, Brazil}

\begin{abstract}
The quantum dynamics of an induced electric dipole in the presence of a configuration of crossed electric and magnetic fields is analyzed. This field configuration confines the dipole in a plane and produces a coupling similar to the coupling of a charged particle in the presence of external magnetic field. In this work we investigate the analog of Landau levels in induced electric dipoles in a sistem of neutral particles. The energy levels and  eigenfunctions are obtained exactly.
\end{abstract}
\maketitle
The quantum dynamics of cold atoms in the presence of an electromagnetic field promises new possibilities to study the quantum properties of many body systems. This system can be easily controlled  and manipulated by electromagnetic fields. The cold atoms make then ideal candidates for the study of several intrinsically quantum phenomena. In recent years the advance of cold atom technology made it possible to simulate  several solid state effects employing neutral atoms and techniques from quantum optics\cite{8,10,9}. Recently, the  neutral atoms techniques to simulate the behavior of a charged particle in this systems\cite{9,11,12,13} has been developted. In this way, the study of systems that simulate the strongly interacting system in a cold atom  has attract much attention in recent years\cite{prl:par,ssc:par}.
 
 Electric dipole moment of  atoms in the presence of an appropriate electromagnetic field configuration has been utilized in literature to inquire above some physical effects\cite{prl:bax,non,any,pac}. The study of quantum phases in the quantum dynamics of the electromagnetic dipole has its origin in the Aharonov-Casher effect\cite{prl:ac} for the magnetic dipole in the presence of an  electric field, that is reciprocal effect of the well known Aharonov-Bohm\cite{pr:ab} effect. The dual effect of Aharonov-Casher effect was studied independently by He and McKellar\cite{pra:mc} and Wilkens\cite{prl:wil}, which  investigated the quantum dynamics of the electric dipole in the presence of magnetic field. In fact, electric dipoles can give rise to a variety of different phenomena that generate geometric phases\cite{pra:mc,prl:wil,spa,pac,euro:furt} in different field-dipole configurations. Wei, Han and Wei\cite{wei}  have proposed a more realistic field configuration that generate  a He-McKellar-Wilkens effect. In fact, all quantum effects in the dynamics of the dipole occur due to the coupling of electric/magnetic dipole with external magnetic/electric   field make  a mimic of minimal coupling of charged particle with an external magnetic field. In this work we use this configuration to study a possibility of an analog Landau quantization due to induced electric dipoles in cold atom systems. 

The interaction of the electromagnetic field with a charged particle, plays an important role in the generation of collective phenomena, such as,  fractional statistics\cite{prange} and the quantum Hall effect.
The quantum motion of a charged particle in the presence of a constant magnetic field is described by the theory of Landau \cite{zlan}. The Landau quantization in two dimensions makes the energy levels coalesce into a discrete spectrum. The Landau levels  present a remarkable interest in several areas of the physics among they the quantum Hall effect\cite{prange}.

 On the order hand, the Landau levels were studied for different curved surfaces\cite{ap1,ap2} with the interest in several areas of physics. The idea of analogs of Landau quantization was proposed initially by Ericsson and Sj\"orvist inspired in the the work of Paredes et al. that studied the possibility of an analog of the  Hall effect  in  Bose-Einstein condensates. The idea of Ericsson and Sj\"oqvist\cite{pra:sjo} was that  the Aharonov-Casher  interaction can be used to generate an analog of Landau levels in systems of neutral atoms. In a recent work we extend this idea to systems of electric dipoles in the presence of a magnetic field and  demonstrated also a similar effect of Landau quantization to electric dipoles\cite{cla:quan}. In the present work we analyze this possibility in a system of induced dipoles in the presence of electric and magnetic fields. The aim this work is to demonstrate that the  Wei, Han Wei configuration for topological phase can be used to investigate the landau quantization in a similar scheme that was treated  in the Landau-Aharonov-Casher effect by  Ericsson and Sj\"oqvist\cite{pra:sjo} and in the Landau-He-McKellar-Wilkens effect by us\cite{cla:quan}. The central idea  is that a system of cold atom in this condition exhibits Landau quantization and can be used in future study of atomic quantum Hall in the condition presented here. The cold atom is treated as structureless induced dipole moment. We consider that our system was submitted the following electric field 
\begin{equation}\label{1}
\mathbf{E}=\frac{\rho}{2} r \hat{e}_{r},
\end{equation} 
where $\rho$ is the charge density. The system is also submitted to the external magnetic field
\begin{equation}\label{2}
\mathbf{B}= B_{0}\hat{e}_{z}.
\end{equation} 
The  presence of  electric field induces an electric dipole in the cold atom given by
\begin{eqnarray}\label{dip}
\mathbf{d}=\alpha(\mathbf{E} + \mathbf{v}\times \mathbf{B})
\end{eqnarray} 
where $\alpha$ is the dielectric polarizability of the cold atoms and $v$ is the velocity of the atom. The Lagrangian that describes  the electric dipole  in the presence of an electromagnetic field is given by
\begin{eqnarray}
L=\frac{1}{2}M\mathbf{v}^2 +  \frac{1}{2} \mathbf{d} \cdot (\mathbf{E} + \mathbf{v}\times \mathbf{B}).
\end{eqnarray} 
Considering that the dipole is induced in cold atoms, the Lagrangian is written as
\begin{center}

\end{center}
\begin{eqnarray}
L=\frac{1}{2}(M + \alpha B^{2})\mathbf{v}^2 +  \frac{1}{2} \alpha E^{2} + \alpha \mathbf{v}\cdot\mathbf{B}\times\mathbf{E}.
\end{eqnarray} 
The term $ \alpha \mathbf{v}\cdot\mathbf{B}\times\mathbf{E}$ is known as the energy of R\"otgen and is responsible for the modification in the canonical momentum of the system that differs from the  mechanical component $Mv$. The study of this term and its similarity with the Chern-Simons term has been widely investigated  recently. Baxter\cite{prl:bax} has showed that the systems of cold Rydberg atoms and Chern-Simons theory are analog to each other. In a similar system,  Zhang\cite{non} has analized the possibility of testing spatial noncommutativity. The study of the duality property has been done in cold Rydberg atoms recently by Noronha and Wotzasek\cite{plb:clov}. The field configuration requires that the system  stays confined in the plane and this  fact is responsible by its analogy with Chern-Simons theory.  
In this way, we can write the  Hamiltonian  associated with this  system as

\begin{equation}\label{effec}
H= \frac{1}{2m^{*}}\left[\vec{P} + \alpha (\vec{E} \times  \vec{B})\right]^{2}-\frac{1}{2}\alpha E^{2},
\end{equation}
where $m^{*}= M + \alpha B^{2}$. Note the similarity of (\ref{effec}) with the Hamiltonian of a charged particle in the presence of an external potential submitted to an external magnetic field. Using this fact the effective  vector potential 
takes this form
\begin{equation}
\mathbf{A}_{eff}=\mathbf{E} \times  \mathbf{B}.
\end{equation} 
Using the field configuration given in (\ref{1}) and (\ref{2}), we obtain the following effective vector potential
\begin{eqnarray}\label{pot}
\mathbf{A}_{eff}= -\frac{B_{0}\rho}{2} r \hat{e}_{\phi}.
\end{eqnarray} 
We define the "magnetic" field $ \mathbf{B}=\bm \nabla \times \mathbf{A}_{eff}$ associated to the effective vector potential as
\begin{equation}
\mathbf{B}_{eff}= B_{0}\rho\hat{e}_{z}.
\end{equation} 
Note that, in this field configuration, the interaction between the electromagnetic field and the electric dipole of atom in the nonrelativistic limit coincides formally with the minimal coupling of a charged particle with an external magnetic field.  (\ref{pot}), in the present case, is responsible to minimal coupling. Similar effects have been pointed out by Ericsson and Sj\"oqvist\cite{pra:sjo} to Landau-Aharonov-Casher levels. In the Ref.\cite{cla:quan} we investigated electric dipole in presence of external magnetic field. In \cite{pra:sjo} and  \cite{cla:quan} the precise configurations for occurrence of a Landau quantization are presented. Note that this field configuration gives the precise condition under which the Landau quantization occurs in the cold atom system. Note also that, the  Wei, Han and  Wei,  field configuration used in this work constrained the motion of the dipole to be planar and to obey these conditions. 
Now, the problem is reduced  to the solution of the Schr\"odinger equation associated to (\ref{effec}).
In this way, we write the Sch\"odinger equation for this system,  in cylindrical coordinates, in the following form
\begin{eqnarray}\label{eqc5.5}
	-\frac{1}{2 m^{*}}\left[\frac{1}{r}\frac{\partial}{\partial r}\left(r\frac{\partial\psi}{\partial r}\right)+\frac{1}{r^2}\frac{\partial^2\psi}{\partial\phi^2}\right]-\\ \nonumber-\frac{i\omega}{2}\frac{\partial\psi}{\partial\phi}+\frac{m^{*} \omega^2}{8}r^2\psi+ \frac{{m^{*}}^{2}\omega^{2}}{8\alpha B_{0}^{2}}r^{2}\psi={\cal{E}}\psi\;,
\end{eqnarray}
where $\omega=\frac{\alpha B_{0} \rho}{m^{*}}$
We use the following Ansatz to the solution of Schr\"odinger equation
\begin{equation}\label{eqc5.7}
	\psi=e^{i \ell \phi}R(r)\;,
\end{equation}
where  $\ell$ is an integer number. Using the Eq. (\ref{eqc5.7}),  the Eq. (\ref{eqc5.5}) assumes the following form:
\begin{eqnarray}\label{eqc5.8}
	\frac{1}{2M}\left(R''+\frac1r R'-\frac{m^2}{r^2}R\right)+ \nonumber \\ + \left({\cal{E}}-\frac{M\omega^2}{8}r^2-\frac{\ell\omega}{2}-\frac{{m^{*}}^{2}\omega^{2}}{8\alpha B_{0}^{2}}r^{2} \right)R=0\;.
\end{eqnarray}
We use the  following change of variables
\begin{eqnarray}
	\xi&=&\frac{m^{*}\omega}{2}r^2\; .
\end{eqnarray}
In this way,  Eq. (\ref{eqc5.8}) is transformed  into
\begin{equation}\label{eqc5.10}
	\xi R''+R'+\left(-\frac{\delta\xi}{4}+\beta-\frac{\ell^2}{4\xi} \right)R=0\;,
\end{equation}
where $\beta=\frac{{\cal{E}}}{\omega} - \frac{m}{2} $ and $\delta=1+\frac{m^{*}}{4\alpha B_{0}^{2}}$.
Assuming for the radial eigenfunction the form
\begin{equation}\label{eq2.67}
	R(\xi)=e^{-\delta \xi/2}\xi^{|\ell|/2}\zeta(\xi)\;,
\end{equation}
which satisfies the usual asymptotic requirements and the finiteness at the origin for the bound state, we have
\begin{equation}\label{eq2.68}
	\xi\frac{d^{2}\zeta}{d\xi^{2}} +\left[(|\ell|+1)-\delta \xi\right]\frac{d\zeta}{d\xi} -\gamma \zeta=0,
\end{equation}
where $\gamma=\beta-\frac{\delta(|\ell|+1)}{2}$. We find that the solution of equation (\ref{eq2.68}) is the degenerated hypergeometric function
 \begin{eqnarray}
\label{hyper}
\zeta(\xi)=F\left[-\gamma,|\ell|+1,\delta \xi\right]\;.
\end{eqnarray} 
In  order to have normalization of the wave function, the series in (\ref{hyper}) must be a polynomial of degree $\nu$, therefore,
\begin{equation}
	\frac{\gamma}{\delta}=\frac{1}{\delta}\beta-\frac{(|\ell|+1)}{2}=n.
\end{equation}
Where $n$ is an integer number. With this condition, we obtain discrete values for the energy, given by
\begin{equation}\label{eq2.69}
	{\cal{E}}_{n,\ell}=\left(n+\frac{|\ell|}{2}+\frac{1}{2}\right)\omega_{\delta} +\frac{\ell}{2}\omega,
\end{equation}
where $\omega_{\delta}=\delta \omega$. In the limit that $\delta\longrightarrow 1$ the eingevalues are given by\begin{eqnarray}
{\cal{E}}_{n,\ell}=\left(n+\frac{|\ell|}{2}+\frac{1}{2} +\frac{\ell}{2} \right)\omega
\end{eqnarray} 
Note that, this limit is characterized by a high magnetic field. In this case the system is similar to Landau levels of a charged particle.
The radial eigenfunction is then given by
\begin{eqnarray}\label{eq2.70}
	R_{\nu,\ell}&=&\frac{1}{a^{1+|\ell|}}\left[\frac{(|\ell|+n)!}{2^{|\ell|}n!|\ell|!^2}\right]^{1/2} \exp\left(-\frac{\delta r^2}{4a^2}\right)\times \nonumber \\ &\times& r^{|\ell|}F\left[-n,|\ell|+1,\frac{\delta r^2}{2a^2}\right]\;,.
\end{eqnarray}

In this work we study the eigenfunctions and the eigenvalues  of a neutral particle(atom)  with an induced dipole moment in the presence of crossed electric and magnetic fields. The field configuration is the same of Wei, Han  and  Wei \cite{wei} and confines particles in two dimensions.  In a strong magnetic field, the energy levels are similar to Landau levels. Based in this fact the possibility of  an atomic analog of the Landau quantization to electric  dipole  is presented in a similar way to Landau quantization investigation of the magnetic dipole\cite{pra:sjo}. This effect can be viewed as a first approach to investigate a atomic analog of quantum Hall effect with electric dipoles in cold atoms. 
\acknowledgments
This work was partially supported by CNPq, CAPES (PROCAD),PRONEX/CNPq/FAPESQ and by CNPq/FINEP .We thank Professor F. Moraes for the critical reading of this manuscript.

\end{document}